\documentstyle[12pt,bezier]{article}
\newfont{\mathea}{msam10 scaled\magstep0}
\newfont{\matheb}{msbm10 scaled 1095}
\newfont{\tmpEins}{cmsy10 scaled 2074}
\newfont{\tmpZwei}{cmsy10 scaled 1095}
\newfont{\tmpDrei}{cmsy10 scaled 1000}
\newfont{\tmpVier}{cmsy5 scaled 1000}
\newfont{\tmpFuenf}{msbm7 scaled\magstep0}

\def\Bbb#1{\mathchoice{\mbox{\matheb #1}}{\mbox{\matheb #1}}%
 {\mbox{\tmpFuenf #1}}{\mbox{\tmpFuenf #1}}}

\def\restriction{\mathchoice{%diplaystyle
 \mbox{\unitlength1cm\begin{picture}(.2,.4)%
  \bezier{5}(.07,.3)(.1,.27)(.13,.24)%
  \put(.07,.35){\line(0,-1){.5}}\end{picture}}}{%textstyle
 \mbox{\unitlength1cm\begin{picture}(.2,.4)%
  \bezier{5}(.07,.3)(.1,.27)(.13,.24)%
  \put(.07,.35){\line(0,-1){.5}}\end{picture}}}{%scriptstyle
 \mbox{\mathea\symbol{22}}}{%scriptscriptstyle
 \mbox{\mathea\symbol{22}}}}

\def\dach#1#2{\mbox{$\mathop{\vbox{\ialign{%
  $##\crcr\hfil #1 \hfil$\crcr}}}\limits^{\scriptscriptstyle #2}$}}

\def\rnzs{\dach{\rho_2}{\mbox{$\scriptscriptstyle\kern-.7mm0$}}\kern-1.2mm'}

\def\Subset{\mbox{$\subset\kern-.5mm\subset$}}
\newcommand{\LI}{\mbox{{\rm L$^{\kern-.15em\raise.2ex\hbox{\scriptsize 1}}$}}}

\def\Ldummy{\left.\bgroup}
\def\Rdummy{\egroup^{\rule{0mm}{1.4mm}}\right.}
\def\LA{\left\langle\bgroup}
\def\RA{\egroup^{\rule{0mm}{1.4mm}}\right\rangle_{\cal A}^{}}
\def\LR{\left(\bgroup}
\def\RR{\egroup^{\rule{0mm}{1.4mm}}\right)}
\def\LG{\left\{\bgroup}
\def\RG{\egroup^{\rule{0mm}{1.4mm}}\right\}}

\def\Wort#1{\mbox{{\rm #1\kern.1em}}}

\def\lfac#1#2{\vcenter{\hbox{$#1\kern-.2em\raise-.6ex\hbox{\Large{/}}%
 \kern-.2em\raise-1.2ex\hbox{$#2$}$}}}

\def\gin{\mbox{\tmpZwei\symbol{91}\kern-1.4mm\rule{.2mm}{1.85mm}\kern1.4mm}}
\def\gni{\mbox{\tmpZwei\symbol{92}\kern-1.4mm\rule[.15mm]{.2mm}{1.85mm}%
  \kern1.4mm}}

\def\EINS{{\mathop{1\kern-.25em\mbox{{\rm{\small l}}}}}}

\begin{document}

\LARGE On Lax-Phillips semigroups 
\normalsize

\vspace{1cm}

H. Baumg\"artel

\vspace{0.3cm}

University of Potsdam, Mathematical Institute

Am Neuen Palais 10, PF 601553

D-14415 Potsdam, Germany

e-mail: baumg@rz.uni-potsdam.de

\begin{abstract}
Lax-Phillips evolutions are described by two-space scattering systems. The canonical
identification operator is characterized for Lax-Phillips evolutions, whose
outgoing and incoming projections commute. In this case a (generalized) 
Lax-Phillips semigroup can be introduced and its spectral theory is considered. In the
special case, originally considered by Lax and Phillips (where the outgoing
and incoming subspaces are mutually orthogonal), this semigroup coincides with that
introduced by Lax and Phillips. In the more general case the existence of the
semigroup is not coupled with the (global) holomorphic continuability of the
scattering matrix into the upper half plane. The basic connection of the Lax-Phillips
semigroup to the so-called characteristic semigroup of the reference evolution
is emphasized.
\end{abstract}

\section{Introduction}
Recently
several papers were published where the mathematical framework of the Lax-Phillips
scattering theory [1] is used
for the description of resonances in quantum theory,
see Strauss [2,3] and papers quoted there, e.g.
Flesia and Piron [4],
Horwitz and Piron [5], Eisenberg and Horwitz [6], Strauss, Horwitz
and Eisenberg [7].
The reason is the existence of a distinguished semigroup
in the Lax-Phillips scattering theory (the Lax-Phillips semigroup)
and the relation between their eigenvalues and poles of the scattering matrix.

A serious obstacle for this point of view is the fact that the Lax-Phillips
evolutions have generators whose spectrum is pure absolutely continuous,
coincides with the real line and has constant multiplicity, whereas Hamiltonians
in quantum mechanics are usually bounded below. However this obstacle can be
overcome, for example by using ideas of
Halmos [8] (refined by Kato [9]). This approach is pointed out in [10].
A further approach is given by Strauss [2] which is based on the theory of
Sz.-Nagy-Foias [11] of contractions operators on Hilbert space.

Therefore it seems to be of interest to pass in review the Lax-Phillips theory
from the pure mathematical point of view with the aim to establish Lax-Phillips 
semigroups under the most general assumptions on the evolution or to replace
its existence by other suitable assumptions.

The results, presented in this paper, suggest to extend the crucial
restriction of the characteristic semigroup (see Subsection 2.3) also for cases
where the semigroup property is violated and to replace this lack by independent 
analyticity assumptions on the scattering matrix. First steps in this direction are 
proposed in [10].

\section{LP-evolutions}

A unitary strongly continuous evolution group
$U(\Bbb{R})$
on a Hilbert space
${\cal H}$
is called an LP-evolution, if there are subspaces
${\cal D}_{+},\,{\cal D}_{-}$ in
${\cal H}$,
called outgoing and incoming, such that
\[
U(t){\cal D}_{+}\subseteq{\cal D}_{+},\,t\geq 0\quad
U(t){\cal D}_{-}\subseteq{\cal D}_{-},\,t\leq 0,
\]
\[
\bigcap_{t\in\Bbb{R}}U(t){\cal D}_{\pm}=\{0\},\quad
\mbox{clo}\{\bigcup_{t\in\Bbb{R}}U(t){\cal D}_{\pm}\}={\cal H}.
\]
These evolutions were introduced by Lax and Phillips in [1], where the basic theorems
are presented and the theory of these evolutions is developed,
especially for the case that outgoing and incoming subspaces are mutually
orthogonal.

\subsection{The reference evolution}

Let
${\cal H}_{0}:=L^{2}(\Bbb{R},dx,{\cal K})$,
where
${\cal K}$
is a separable Hilbert space and
\[
T(t)f(x):= f(x-t),\quad f\in {\cal H}_{0}
\]
the regular translation group representation on
${\cal H}_{0}$,
(where multiplicity $\dim{\cal K}$ is taken into account).

\vspace{0.3cm}

For convenience of the reader we recall the properties of this 
LP-evolution (see e.g. [12, p. 250 ff.]):
\begin{equation}
(P_{\pm}f)(x):=\chi_{\Bbb{R}_{\pm}}(x)f(x),\quad f\in{\cal H}_{0},
\end{equation}
where
$\Bbb{R}_{+}:=[0,\infty),\;\Bbb{R}_{-}:=(-\infty,0]$,
are the projections onto the outgoing/incoming subspaces.
\[
P_{\pm}(t):=T(-t)P_{\pm}T(t),\quad t\in\Bbb{R}.
\]
The function
$t\rightarrow P_{+}(t)$
is monotonically increasing,
\[
P_{+}(t_{1})\leq P_{+}(t_{2}),\quad t_{1}\leq t_{2},
\]
and
\[
\mbox{s-}\lim\limits_{t\rightarrow +\infty}\,P_{+}(t)=
\EINS_{{\cal H}_{0}},\quad
\mbox{s-}\lim\limits_{t\rightarrow -\infty}\,P_{+}(t)=0.
\]
Similarly,
$P_{-}(\cdot)$
is monotonically decreasing and
\begin{equation}
\mbox{s-}\lim\limits_{t\rightarrow+\infty}\,P_{-}(t)=0,\quad 
\mbox{s-}\lim\limits_{t\rightarrow-\infty}P_{-}(t)=\EINS_{{\cal H}_{0}}.
\end{equation}
Furthermore,
$T(t)P_{+}{\cal H}_{0}\subseteq P_{+}{\cal H}_{0}$
for
$t\geq 0$
or
\[
T(t)P_{+}=P_{+}T(t)P_{+},\quad t\geq 0,
\]
correspondingly
\[
T(t)P_{-}=P_{-}T(t)P_{-},\quad t\leq 0.
\]
The unitary evolution group
$T(\cdot)$
on
${\cal H}_{0}$
is called the {\em reference} LP-evolution,
$P_{+}{\cal H}_{0}$
is the {\em outgoing} and
$P_{-}{\cal H}_{0}$
the {\em incoming} subspace. In this case
$P_{+}{\cal H}_{0}$
and
$P_{-}{\cal H}_{0}$
are mutually orthogonal and
$P_{+}{\cal H}_{0}\oplus P_{-}{\cal H}_{0}={\cal H}_{0}.$

\vspace{0.3cm}

By Fourier transformation
the representation
$T(\Bbb{R})$
is transformed into
\[
\hat{T}(t):=FT(t)F^{-1},
\]
where
\[
(\hat{T}(t)\hat{f})(p)=e^{-itp}\hat{f}(p),\quad \hat{f}\in{\cal H}_{0},
\]
i.e. the multiplication operator
$H_{0}$
on
${\cal H}_{0}$
given by
\[
(H_{0}\hat{f})(p):=p\hat{f}(p),\quad \hat{f}\in\mbox{dom}\,H_{0},
\]
is the generator of
$\hat{T}(\Bbb{R}):$
\[
\hat{T}(t)=e^{-itH_{0}},\quad t\in\Bbb{R}.
\]
$\hat{T}(\Bbb{R})$
is called the {\em spectral representation} of the reference evolution.
One has
$\mbox{spec}\,H_{0}=\Bbb{R}$
and it is pure absolutely continuous. 
Note that we use the Fourier transformation in the form
\[
(Ff)(p):=(2\pi)^{-1/2}\int_{-\infty}^{\infty}e^{-ipx}f(x)dx
\]

\vspace{0.3cm}

The projection
$P_{+}$,
defined by (1), is an element from the spectral measure of
$H_{0}$,
therefore
$P_{+}\hat{T}(t)=\hat{T}(t)P_{+},\;t\in\Bbb{R}$
and
$\hat{T}(t)\restriction P_{+}{\cal H}_{0}$
is a positive representation,
$\mbox{spec}\,(H_{0}\restriction P_{+}{\cal H}_{0})=[0,\infty)$,
and it is pure absolutely continuous.
The projections
\[
Q_{\mp}:=FP_{\pm}F^{-1}
\]
are the projections onto the Hardy spaces
${\cal H}^{2}_{\mp}(\Bbb{R},{\cal K})=:{\cal H}^{2}_{\mp}\subset{\cal H}_{0}$
(see e.g. [13]). That is, these spaces are outgoing/incoming subspaces for
$\hat{T}(\Bbb{R})$
and
$Q_{\pm}$
are the corresponding projections. 

The projection
$Q_{+}$
is given by
\begin{equation}
{\cal H}_{0}\ni g\rightarrow (Q_{+}g)(z)=(2i\pi)^{-1}
\int_{-\infty}^{\infty}\frac{g(\lambda)}{\lambda-z}d\lambda.
\end{equation}

\subsection{The main theorem for LP-evolutions}

Let
$U(\Bbb{R})$
be an LP-evolution on
${\cal H}$
with outgoing/incoming subspaces
${\cal D}_{\pm}$.
Then there are isometric operators
$V_{\pm}$
from
${\cal H}$
onto
${\cal H}_{0}$ 
with an appropriate multiplicity space
${\cal K}$
such that
\[
V_{\pm}U(t)V_{\pm}^{\ast}=e^{-itH_{0}},\quad t\in\Bbb{R}
\]
and
\[
Q_{\mp}{\cal H}_{0}=V_{\pm}{\cal D}_{\pm}.
\]
The isometries $V_{\pm}$ are unique up to isomorphisms of ${\cal
K}$. This means, if $V_{\pm}'$ is a second pair of isometries
then there are unitaries
$K_{\pm}$
on
${\cal K}$
such that 
$V_{+}'=K_{+}V_{+},\;V_{-}'=K_{-}V_{-}$
where
$(K_{\pm}f)(\lambda):=K_{\pm}f(\lambda)$
(see Sinai [14] and Lax and Phillips [1], see also [12]).
$V_{\pm}$
maps onto the so-called {\em outgoing/incoming spectral representation} of
$U(\Bbb{R})$.
In general
$V_{+}\neq V_{-}$.

An important implication of the main theorem is that
$U(t)=e^{-itH}$,
where
$\mbox{spec}\,H=\Bbb{R}$
and $H$ has constant multiplicity.

We introduce the
orthoprojectios 
$D_{\pm}$
onto the subspaces 
${\cal D}_{\pm}$.
Then
\[
D_{+}=V_{+}^{\ast}Q_{-}V_{+},\quad
D_{-}=V_{-}^{\ast}Q_{+}V_{-}
\]
and
${\cal D}_{+}=V_{+}^{\ast}{\cal H}^{2}_{-},\,
{\cal D}_{-}=
V_{-}^{\ast}{\cal H}
^{2}_{+}.$

The LP-scattering operator is defined by
$S_{LP}:=V_{+}V_{-}^{-1}.
\;S
_{LP}$
commutes with the reference evolution, i.e.
\[
S_{LP}e^{-itH_{0}}=e^{-itH_{0}}S_{LP},
\]
therefore
$S_{LP}$
acts as
\[
(S_{LP}f)(\lambda)=S_{LP}(\lambda)f(\lambda),\quad
f\in {\cal H}_{0}.
\]
The operators
$S_{LP}(\lambda)$
are unitaries on
${\cal K}$
a.e. on
$\Bbb{R}$.
The operator function
$S_{LP}(\cdot)$
is called the LP-scattering matrix.

\subsection{Semigroups connected with the reference evolution}

First the semigroup
\begin{equation}
T_{+}(t):=Q_{+}e^{-itH_{0}}Q_{+}=Q_{+}e^{-itH_{0}},\quad t\geq 0,
\end{equation}
is considered, resp. its restriction
$T_{+}(t)\restriction {\cal H}^{2}_{+}$,
which we call the
{\em characteristic semigroup}.
It plays an important role as as "intermediate step" to obtain
the Lax-Phillips semigroup. It was already introduced by Y. Strauss [3].
Further we need its adjoint
\begin{equation}
T_{+}(t)^{\ast}=Q_{+}e^{itH_{0}}Q_{+}=e^{itH_{0}}Q_{+},\quad t\geq 0,
\end{equation}
resp. 
$T_{+}(t)^{\ast}\restriction {\cal H}^{2}_{+}$.
The last equations in(4) and (5) are true because
$Q_{+}$
is the incoming projection for
$\hat{T}(\cdot)$,
i.e. it is the outgoing projection for
$\hat{T}(\cdot)^{\ast}$.

First we recall the properties of
$T_{+}(\cdot)^{\ast}\restriction {\cal H}^{2}_{+}$.
It is a strongly continuous and isometric semigroup, i.e.
\[
\Vert T_{+}(t)^{\ast}f\Vert=\Vert f\Vert,\quad f\in{\cal H}_{+}^{2},
\]
we have
\[
T_{+}(t)^{\ast}\restriction{\cal H}^{2}_{+}=e^{itC_{-}},\quad t\geq 0,
\]
and the generator
$C_{-}$,
a closed operator on
${\cal H}^{2}_{+}$,
with  domain
$\mbox{dom}\,C_{-}$
dense in
${\cal H}^{2}_{+}$,
satisfies
\begin{equation}
\Bbb{C}_{-}\subset\mbox{res}\,C_{-},
\end{equation}
where
$\Bbb{C}_{-}:=\{\zeta\in\Bbb{C}:\mbox{Im}\,\zeta<0\}$.

\vspace{3mm}

PROPOSITION 1. {\em The generator}
$C_{-}$
{\em satisfies the following properties:}
\begin{itemize}
\item[(i)]
$\mbox{dom}\,C_{-}=\{f\in\mbox{dom}\,H_{0}\cap{\cal H}^{2}_{+}:
H_{0}f\in{\cal H}^{2}_{+}\}$
{\em and}
\[
(C_{-}f)(z)=zf(z),\quad \mbox{Im}\,z>0,\quad f\in\mbox{dom}\,C_{-},
\]
\item[(ii)]
{\em the deficiency space}
\[
{\cal N}_{\zeta}:={\cal H}^{2}_{+}\ominus (\zeta-C_{-})\mbox{dom}\,C_{-},\quad
\mbox{Im}\,\zeta>0
\]
{\em is given by}
\begin{equation}
{\cal N}_{\zeta}=\{f\in{\cal H}^{2}_{+}: f(z)=(z-\overline{\zeta})^{-1}k,\quad
k\in{\cal K}\}.
\end{equation}
{\em Moreover,}
$(\zeta-C_{-})\mbox{dom}\,C_{-}$
{\em is a subspace and it coincides with}
\[
{\cal M}_{\zeta}:=\{f\in{\cal H}^{2}_{+}: f(\zeta)=0\}.
\]
\end{itemize}

\vspace{3mm}

Proof. (i) is obvious because of (5).
(ii) First we prove that 
${\cal M}_{\zeta}$
is a subspace. Let
$f_{n}\in {\cal H}^{2}_{+},\;f_{n}(\zeta)=0$
and
$\Vert f_{n}-f\Vert\rightarrow 0$
for
$n\rightarrow\infty$,
where
$f\in {\cal H}^{2}_{+}.$
We have to show that
$f(\zeta)=0.$
We put
\[
h_{\zeta}(x):=\frac{1}{x-\zeta}.
\]
Then
$h_{\zeta}\in L^{2}(\Bbb{R},dx)$.
According to (3) we have
\[
f(\zeta)=\frac{1}{2\pi i}\int_{-\infty}^{\infty}h_{\zeta}(x)f(x)dx,\quad
f_{n}(\zeta)=\frac{1}{2\pi i}\int_{-\infty}^{\infty}h_{\zeta}(x)
f_{n}(x)dx.
\]
Then
\[
\Vert f(\zeta)-f_{n}(\zeta)\Vert_{\cal K}\leq
\frac{1}{2\pi}\int_{-\infty}^{\infty}\vert h_{\zeta}(x)\vert\cdot
\Vert f(x)-f_{n}(x)\Vert_{\cal K}dx\leq
\]
\[
\frac{1}{2\pi}\Big(\int_{-\infty}^{\infty}\vert h_{\zeta}(x)\vert^{2}dx
\Big)^{1/2}\cdot\Big(\int_{-\infty}^{\infty}\Vert f(x)-f_{n}(x)\Vert
_{\cal K}^{2}dx\Big)^{1/2}.
\]
This implies
$\Vert f(\zeta)-f_{n}(\zeta)\Vert_{\cal K}\rightarrow 0$
hence
$f(\zeta)=0$
follows. Now we prove
$(\zeta-C_{-})\mbox{dom}\,C_{-}={\cal M}_{\zeta}$.
The inclusion
$\subseteq$
is obvious because for
$f\in\mbox{dom}\,C_{-}$
the function
$g(z):=(\zeta-z)f(z)$
vanishes at the point
$\zeta$
i.e.
$g(\zeta)=0.$
To prove the other inclusion let
$f\in{\cal M}_{\zeta}$,
i.e.
$f(\zeta)=0$.
Then
\begin{equation}
f(z)=(z-\zeta)g(z),
\end{equation}
where the function
\[
g(z):=\frac{f(z)}{z-\zeta}
\]
is holomorphic on the upper half plane. Moreover, one calculates
easily that
$g\in {\cal H}^{2}_{+}.$
Now from (8) one gets
\[
zg(z)=\zeta g(z)+f(z)
\]
and the right hand side is an element of
${\cal H}^{2}_{+}$.
Therefore
$g\in\mbox{dom}\,C_{-}$
follows, i.e.
$f\in (\zeta-C_{-})\mbox{dom}\,C_{-}.$

Finally we prove (7):
Let
\[
f_{\overline{\zeta},k}(z):=\frac{k}{z-\overline{\zeta}},\,k\in{\cal K}\quad
\mbox{and}\quad g\in {\cal H}^{2}_{+}.
\]
Then
\begin{equation}
(f_{\overline{\zeta},k},g)=\int_{-\infty}^{\infty}\Big(\frac{k}{x-\overline{\zeta}
},g(x)\Big)_{\cal K}dx=
\int_{-\infty}^{\infty}\frac{1}{x-\zeta}
(k,g(x))_{\cal K}dx=2i\pi(k,g(\zeta))_{\cal K}.
\end{equation}
Now, if
$g\in{\cal M}_{\zeta}$
then
$f_{\overline{\zeta},k}\bot g$
follows or
$f_{\overline{\zeta},k}\in{\cal M}_{\zeta}^{\bot}.$
On the other hand, if
$(f_{\overline{\zeta},k},g)=0$
for all
$k\in{\cal K}$
then
$(k,g(\zeta))_{\cal K}=0$
follows, i.e.
$g(\zeta)=0$
or
$g\in{\cal M}_{\zeta}.\quad \Box$

\vspace{0.3cm}

Proposition 1 implies that the deficiency number 
$\dim {\cal N}_{\zeta}$
of $C_{-}$ w.r.t. the upper half plane coincides with
$\dim{\cal K}.$
(6) implies that
the deficiency number of $C_{-}$ for the lower half plane is 0.

$C_{-}$ is even maximal symmetric, there is no symmetric
extension of $C_{-}$.

Now let
$C_{-}^{\ast}$
be the adjoint of $C_{-}$. Then $C_{-}^{\ast}$ is an extension
of $C_{-},\; C_{-}\subset C_{-}^{\ast}.$

\vspace{3mm}

PROPOSITION 2. {\em The adjoint}
$C_{-}^{\ast}${\em of}
$C_{-}$
{\em satisfies the following properties}:
\begin{itemize}
\item[(i)]
{\em One has}
\[
\mbox{dom}\,C_{-}^{\ast}=\mbox{dom}\,C_{-}\oplus{\cal N}_{\overline{\zeta}},
\]
{\em where}
$\mbox{Im}\,\zeta<0,\,\zeta$
{\em fixed but arbitrary and}
\[
C_{-}^{\ast}f=\zeta f,\quad f\in{\cal N}_{\overline{\zeta}},
\]
{\em i.e. each point}
$\zeta\in\Bbb{C}_{-}$
{\em is an eigenvalue of}
$C_{-}^{\ast}$
{\em and the corresponding eigenspace is given by}
${\cal N}_{\overline{\zeta}}$
{\em i.e all eigenvectors are given by}
\[
\Bbb{C}_{+}\ni z\rightarrow f_{\zeta,k}(z)
:=\frac{k}{z-\zeta},\quad
k\in{\cal K},\quad \mbox{Im}\,\zeta<0.
\]
\item[(ii)]
$\frac{1}{2i\pi}f_{\zeta,k}$
{\em coincides with the Dirac linear forms (evaluation forms) for the
scalar holomorphic function}
$\Bbb{C}_{+}\ni z\rightarrow (k,f(z))_{\cal K}$
{\em on the upper half plane.}
\end{itemize}

\vspace{0.3cm}

Proof .(i) is obvious because of the formulas of v.Neumann (see for example [15, p.292]).
(ii) follows from the "boundary value formula" (9) for Hardy class functions.$\,\Box$

\vspace{0.3cm}

Concerning the semigroup (4) we obtain

\vspace{3mm}

PROPOSITION 3. {\em The semigroup}
$t\rightarrow T_{+}(t)\restriction{\cal H}^{2}_{+}$
{\em has the following properties}:
\begin{itemize}
\item[(i)]
{\em It is strongly continuous and contractive, i.e.}
\[
T_{+}(t)\restriction{\cal H}^{2}_{+}=e^{-itC_{+}},\quad t\geq 0,
\]
{\em where the generator}
$C_{+}$
{\em is closed on}
${\cal H}^{2}_{+},\;\mbox{dom}\,C_{+}$
{\em is dense and}
$\Bbb{C}_{+}\subset\mbox{res}\,C_{+}.$
\item[(ii)]
\[
C_{+}=C_{-}^{\ast}.
\]
\item[(iii)]
\[
(T_{+}(t)f)(z)=\frac{1}{2i\pi}\int_{-\infty}^{\infty}
\frac{e^{-it\lambda}}{\lambda-z}f(\lambda)d\lambda,\quad f\in {\cal H}^{2}_{+}.
\]
\item[(iv)]
{\em One has}
\[
\mbox{s-lim}_{t\rightarrow\infty}e^{-itC_{+}}=0.
\]
\end{itemize}

\vspace{0.3cm}

Proof. (i) is obvious. (ii) One has
\[
\int_{0}^{\infty}e^{itz}e^{-itC_{+}}dt=i(z-C_{+})^{-1},\quad z\in\Bbb{C}_{+}.
\]
Then
\[
\int_{0}^{\infty}e^{-it\overline{z}}(e^{-itC_{+}})^{\ast}dt=
-i((z-C_{+})^{-1})^{\ast}=-i((z-C_{+})^{\ast})^{-1}=-i(\overline{z}-C_{+}^{\ast})^{-1}.
\]
On the other hand the left hand side equals
\[
\int_{0}^{\infty}e^{-it\overline{z}}e^{itC_{-}}dt=-i(\overline{z}-C_{-})^{-1},
\]
hence
$(\overline{z}-C_{+}^{\ast})^{-1}=(\overline{z}-C_{-})^{-1}$
follows for all
$\overline{z}\in\Bbb{C}_{-}$.
This implies the assertion.

(iii) follows from (3). (iv) One has
\[
T_{+}(t)^{\ast}T_{+}(t)=e^{itH_{0}}Q_{+}e^{-itH_{0}}=
F(T(-t)P_{-}T(t))F^{-1}=FP_{-}(t)F^{-1}
\]
which, according to (2), converges strongly to zero for
$t\rightarrow\infty$,
i.e. one has
\[
\mbox{s-lim}_{t\rightarrow\infty}T_{+}(t)^{\ast}T_{+}(t)
\restriction{\cal H}^{2}_{+}=0.
\]
However
$T_{+}(t)^{\ast}\restriction{\cal H}^{2}_{+}$
is isometric, therefore
$\mbox{s-}\lim_{t\rightarrow\infty}e^{-itC_{+}}=0$
follows.
$\,\Box$

\vspace{3mm}

PROPOSITION 4. {\em Let}
$T_{+}(t)\restriction{\cal H}^{2}_{+}=Q_{+}e^{-itH_{0}}\restriction{\cal H}^{2}_{+},\,
t\geq 0,$
{\em as before. Then}
\begin{itemize}
\item[(i)]
$\mbox{res}\,C_{+}=\Bbb{C}_{+}.$
\item[(ii)]
{\em The eigenvalue spectrum of}
$C_{+}$
{\em coincides with}
$\Bbb{C}_{-}$,
{\em i.e. a real point cannot be an eigenvalue}.
\item[(iii)]
{\em The eigenspace of the eigenvalue}
$\zeta\in\Bbb{C}_{-}$
{\em is given by the following subspace}
\[
{\cal N}_{\overline{\zeta}}:=\{f\in{\cal H}^{2}_{+}: f(z)
:=\frac{k}{z-\zeta},\,k\in{\cal K}\}.
\]
{\em Then}
\begin{equation}
T_{+}(t)f=e^{-it\zeta}f,\quad f\in{\cal N}_{\zeta}
\end{equation}
{\em follows}.
\end{itemize}

\vspace{2mm}

Proof. It is obvious because of Proposition 3. The equations

\begin{eqnarray*}
(T_{+}(t)f_{\zeta,k},g) &=&
(f_{\zeta,k},T_{+}(t)^{\ast}g) \\
&=& 2i\pi(k,e^{it\overline{\zeta}}g(\overline{\zeta}))_{\cal K} \\
&=& 2i\pi e^{it\overline{\zeta}}(k,g(\overline{\zeta}))_{\cal K} \\
&=& e{it\overline{\zeta}}(f_{\zeta,k},g) \\
&=& (e^{-it\zeta}f_{\zeta,k},g)
\end{eqnarray*}
for
$g\in{\cal H}^{2}_{+}$
and
$f_{\zeta,k}(z)=\frac{k}{z-\zeta}$
proves relation (10) directly. $\quad \Box$

\vspace{2mm}

The v.Neumann characterization of
$\mbox{dom}\,C_{+}$
can be rewritten into the following modified one.

\vspace{3mm}

PROPOSITION 5. $f\in\mbox{dom}\,C_{+}$ {\em iff the function}
\[
g_{f}(z):=zf(z)-\frac{i}{\sqrt{2\pi}}\lim_{x\rightarrow -0}(F^{-1}f)(x)
\]
{\em is from}
${\cal H}^{2}_{+}$.
{\em Then}
$C_{+}f=g_{f}.$

\vspace{2mm}

Proof. Without restriction of generality one can choose
$\zeta:=-i$
as the reference point of the v.Neumann characterization. (i) Let
$f(z):=a(z)+\frac{k}{i+z},\,k\in{\cal K}.$
Then
\[
g_{f}(z)=za(z)+k(1-\frac{i}{i+z})-\frac{i}{\sqrt{2\pi}}\lim_{x\rightarrow -0}
(F^{-1}a(x)+kF^{-1}\{(i+z)^{-1}\}(x))
\]
Using
\[
\frac{i}{\sqrt{2\pi}}\lim_{x\rightarrow -0}F^{-1}\{(i+z)^{-1}\}(x)=1,\quad
\lim_{x\rightarrow -0}(F^{-1}a)(x)=0,
\]
one obtains
$g_{f}\in{\cal H}^{2}_{+}$.
(ii) Conversely, let
$f\in{\cal H}^{2}_{+}$
and
$g_{f}\in{\cal H}^{2}_{+}$.
The last term in the expression for
$g_{f}$
is a constant
$k\in{\cal K}$,
i.e. we have
$z\rightarrow zf(z)-k$
is from
${\cal H}^{2}_{+}$.
Now
$z\rightarrow b(z):=\frac{k}{z+i}$
is from
${\cal H}^{2}_{+}$,
hence also
$z\rightarrow z(f(z)-\frac{k}{z+i})$
is from
${\cal H}^{2}_{+}$,
i.e. the functions
$z\rightarrow a(z):=f(z)-\frac{k}{z+i}$
and
$z\rightarrow za(z)$
are from
${\cal H}^{2}_{+}$,
i.e.
$f=a+b$,
where
$a\in\mbox{dom}\,C_{-}$
and
$b\in{\cal N}_{i}.\,\Box$

\subsection{Two-space scattering} 

There is a one-to-one correspondence between LP-evolutions and
complete two-space scattering systems
$\{H,H_{0}\}$,
whose identification operators satisfy
characteristic conditions.
$H_{0}$
denotes, as before, the generator of the reference
LP-evolution.

Let
${\cal H}$
be a  Hilbert space and
$\Bbb{R}\ni t\rightarrow U(t)=e^{-itH}$
a strongly continuous unitary group on
${\cal H}$.
Further let
${\cal H}_{0}$
be as before and
\[
J:\,{\cal H}_{0}\rightarrow{\cal H}
\]
a bounded linear operator. Then one can consider the
two-space wave operators
\[
W_{\pm}:=\mbox{s-lim}_{t\rightarrow\pm\infty}\,U(-t)Je^{-itH_{0}},\quad
\]
(see e.g. [12, p. 168 ff.]). Usually $J$ is called the {\em
identification operator}.

Since the aim is to reformulate LP-scattering in the framework of two-space scattering
w.r.t
${\cal H}_{0}$ and ${\cal H}$
we assume a priori that the wave operators
$W_{\pm}:{\cal H}_{0}\rightarrow{\cal H}$
are isometric, i.e.
$W_{\pm}^{\ast}W_{\pm}=\EINS_{{\cal H}_{0}}$
and also 
{\em complete}, i.e. 
$W_{\pm}W_{\pm}^{\ast}=\EINS_{\cal H}$.
The scattering operator $S$ is given by
$S:=W_{+}^{\ast}W_{-}$.

\vspace{0.1cm}

Two (identification) operators 
$J,\tilde{J}$
are called asypmptotically equivalent if
$W_{\pm}(J)=W_{\pm}(\tilde{J}).$
This condition is equivalent to
\[
\Vert (J-\tilde{J})e^{-itH_{0}}f\Vert\rightarrow 0,\quad t\rightarrow\pm\infty
\]
for all
$f\in{\cal H}_{0}$. Now it is always possible to replace 
$J$ by an equivalent identification operator 
$\tilde{J}$
such that
\begin{equation}
W_{\pm}Q_{\mp}=\tilde{J}Q_{\mp}.
\end{equation}
We put
\begin{equation}
\tilde{J}:=W_{+}Q_{-}+W_{-}Q_{+}.
\end{equation}
Then one calculates easily
$W_{\pm}(\tilde{J})=W_{\pm}(J)$
and (11). That is, for our purpose without restriction of generality we may
assume that the identification operator
$J$
is given by (12).
It is called the {\em canonical} identification operator.
This identification operator satisfies the equations
\begin{equation}
J^{\ast}J=\EINS_{{\cal H}_{0}}+Q_{+}S^{\ast}Q_{-}+Q_{-}SQ_{+}
\end{equation}
and
\begin{equation}
JJ^{\ast}=W_{+}Q_{-}W_{+}^{\ast}+W_{-}Q_{+}W_{-}^{\ast}.
\end{equation}
Note that
$W_{+}Q_{-}W_{+}^{\ast},\,W_{-}Q_{+}W_{-}^{\ast}$
are projections which do not commute in general. These equations lead to

\vspace{3mm}

LEMMA 1. $J^{\ast}J$ {\em is asymptotically equivalent to}
$\EINS_{{\cal H}_{0}}$,
{\em i.e.}
$J^{\ast}$
{\em is an asymptotic left inverse for}
$J$,
{\em and}
$JJ^{\ast}$
{\em is asymptotically equivalent to}
$\EINS_{\cal H}$, {\em i.e.}
$J$
{\em is an asymptotic left inverse for}
$J^{\ast}$.

\vspace{1mm}

Proof. One has
\[
e^{itH_{0}}(J^{\ast}J-\EINS_{{\cal H}_{0}})e^{-itH_{0}}=
\]
\[
e^{itH_{0}}Q_{+}e^{-itH_{0}}S^{\ast}e^{itH_{0}}Q_{-}e^{-itH_{0}}+
e^{itH_{0}}Q_{-}e^{-itH_{0}}Se^{itH_{0}}Q_{+}e^{-itH_{0}},
\]
hence
\[
\mbox{s-lim}_{t\rightarrow\pm\infty} (J^{\ast}J-\EINS_{{\cal H}_{0}})e^{-itH_{0}}
\rightarrow 0,\quad t\rightarrow\pm\infty
\]
follows. Similarly for the second property.

\subsection{LP-evolutions as two-space scattering systems}

Let
$U(\Bbb{R})$
be an LP-evolution on
${\cal H},\,{\cal D}_{\pm}$
the outgoing/incoming subspaces,
$V_{\pm}$
the isometric operators from
${\cal H}$
onto
${\cal H}_{0}$
(with an appropriate multiplicity space
${\cal K}$) such that
$V_{\pm}U(t)V_{\pm}^{\ast}=e^{-itH_{0}}$.
Then one has

\vspace{3mm}

PROPOSITION 6. {\em Let}
$U(\Bbb{R}),{\cal H},{\cal H}_{0},{\cal D}_{\pm},V_{\pm}$
{\em as above.Put}
\[
J:= V_{+}^{\ast}Q_{-}+V_{-}^{\ast}Q_{+}.
\]
{\em Then}
\[
U(t)JQ_{-}=Je^{-itH_{0}}Q_{-},\quad t\geq 0,\quad
U(t)JQ_{+}=Je^{-itH_{0}}Q_{+},\quad t\leq 0,
\]
{\em and the two-space wave operators exist and are given by}
\[
W_{+}=V_{+}^{\ast},\quad W_{-}=V_{-}^{\ast},
\]
{\em i.e. they are isometric and complete. That is: w.r.t.}
$J$
{\em the given LP-evolution}
$U(\Bbb{R})$ 
{\em forms, together with the reference evolution, a complete 
two-space scattering system and its scattering operator}
$S$
{\em coincides with the LP-scattering operator}
$S_{LP}$.

\vspace{1mm}

The proof is given by straightforward calculation (see e.g. [12, p.255 ff.], where 
only the case
${\cal D}_{+}\bot{\cal D}_{-}$
is considered). Conversely, one has

\vspace{3mm}

PROPOSITION 7. {\em Let}
$\{H,H_{0};J\}$
{\em be a complete two-space scattering system with (isometric)  wave operators}
$W_{\pm}$,
{\em such that}
$J$
{\em can be given by}
\begin{equation}
J:=W_{+}Q_{-}+W_{-}Q_{+}.
\end{equation}
{\em Then}
$\{U(\Bbb{R}),{\cal D}_{\pm}\}$,
{\em where}
$U(t):=e^{-itH}$,
{\em is an LP-evolution where the outgoing/incoming subspaces are given by}
${\cal D}_{+}:=W_{+}{\cal H}^{2}_{-},\,{\cal D}_{-}:=W_{-}{\cal H}^{2}_{+}$,
{\em i.e. their projections by}
\begin{equation}
D_{+}:=W_{+}Q_{-}W_{+}^{\ast}=JQ_{-}J^{\ast},\quad
D_{-}:=W_{-}Q_{+}W_{-}^{\ast}=JQ_{+}J^{\ast}.
\end{equation}
{\em The corresponding transformations to the out/in spectral representations
are given by}
$V_{+}:=W_{+}^{\ast},\,V_{-}:=W_{-}^{\ast}$.
{\em The LP-scattering operator}
$S_{LP}$
{\em and}
$S$
{\em coincide}.

\vspace{1mm}

Proof. The equation (15)
implies
\[
e^{-itH}JQ_{-}=Je^{-itH_{0}}Q_{-},\quad t\geq 0,\quad
e^{-itH}JQ_{+}=Je^{-itH_{0}}Q_{+},\quad t\leq 0.
\]
and the equations in (16). Further, the equation
\[
U(-t)D_{+}U(t)=U(-t)W_{+}Q_{-}W_{+}^{\ast}U(t)=W_{+}e^{itH_{0}}Q_{-}
e^{-itH_{0}}W_{+}^{\ast},\quad t\in\Bbb{R}
\]
shows that
$D_{+}$
is an outgoing projection w.r.t.
$U(\cdot)$.
Similarly for
$D_{-}.\,\Box$

\section{Lax-Phillips evolutions with commuting outgoing/incoming projections}
\subsection{Identification operators}
Let
$\{H,H_{0};J\}$ and the associated LP-evolution
$\{U(\Bbb{R}),{\cal D}_{\pm}\}$
be as in Proposition 7, in particular
$J$
is given by formula (15).
Then the question arises in which case
$D_{+}$
and
$D_{-}$
commute,
$D_{+}D_{-}=D_{-}D_{+}.$
First we consider the special case that
$D_{+}D_{-}=0$,
i.e.
${\cal D}_{+}$
and
${\cal D}_{-}$
are mutually orthogonal.

 In this case Lax and Phillips introduced in [1, Chap. III] their famous 
semigroup, which is a special restriction of the semigroup (4)
in Subsection 2.3.

Later on we show that also in the case of commuting projections
$D_{+},\,D_{-}$
the corresponding restriction leads to a semigroup (see Subsection 3.2).

\vspace{0.3cm}

PROPOSITION 8. {\em Let}
$\{U(\Bbb{R}),{\cal D}_{\pm}\}$
{\em be as before. Then the following conditions are equivalent}:
\begin{itemize}
\item[(i)]
$J$ {\em is isometric},
\item[(ii)]
${\cal D}_{+}\bot{\cal D}_{-}$,
\item[(iii)]
$SQ_{+}=Q_{+}SQ_{+}.$
\end{itemize}

\vspace{0.3cm}

Proof.(i)$\leftrightarrow$ (iii): One calculates
\[
J^{\ast}J=(W_{+}Q_{-}+W_{-}Q_{+})^{\ast}(W_{+}Q_{-}+W_{-}Q_{+})=
Q_{-}+Q_{+}S^{\ast}Q_{-}+Q_{-}SQ_{+}+Q_{+}.
\]
If
$J^{\ast}J=\EINS_{{\cal H}_{0}}$
then
$Q_{+}S^{\ast}Q_{-}+Q_{-}SQ_{+}=0$
follows, i.e.
$Q_{-}SQ_{+}=0$
or (iii) and vice versa.

(ii)$\leftrightarrow$ (iii): Using
$D_{\pm}=V^{\ast}_{\pm}Q_{\mp}V_{\pm}$
one obtains
\[
D_{+}D_{-}=V_{+}^{\ast}Q_{-}V_{+}V_{-}^{\ast}Q_{+}V_{-}
=V_{+}^{\ast}Q_{-}SQ{+}V_{-}
\]
and the assertion is obvious.
$\,\Box$

\vspace{2mm}

The characterization of
$J$
in the general case (commuting outgoing and incoming projections) is given by

\vspace{3mm}

THEOREM 1. {\em Let}
$\{U(\Bbb{R}),{\cal D}_{\pm}\}$
{\em be as before. Then}
\[
D_{+}D_{-}=D_{-}D_{+}\quad \mbox{iff}\quad J^{\ast}J=\EINS_{{\cal H}_{0}}
+E-F,
\]
{\em where}
$E,\,F$
{\em are selfadjoint projections with}
$EF=0$.

{\em Moreover either}
$E=F=0$
{\em or both projections are nonzero,}
$E\neq 0,\,F\neq 0.$

\vspace{3mm}

Note that the first case of the last statement corresponds to
${\cal D}_{+}\bot{\cal D}_{-},$
the second one to
$D_{+}D_{-}\neq 0.$

\vspace{2mm}

Proof. (i) Assume
$D_{+}D_{-}=D_{-}D_{+}$.
Then a straightforward calculation yields that this is equivalent to
\begin{equation}
Q_{-}SQ_{+}S^{\ast}=SQ_{+}S^{\ast}Q_{-}.
\end{equation}
Using (13) we have
$J^{\ast}J=\EINS_{{\cal H}_{0}}+A+A^{\ast}$,
where
$A:=Q_{+}S^{\ast}Q_{-}$.
That is, we have to prove
$A+A^{\ast}=E-F$,
where
$E,\,F$
have the mentioned properties. Note that
\[
(A+A^{\ast})^{2}=AA^{\ast}+A^{\ast}A,\quad
(AA^{\ast}+A^{\ast}A)^{2}=AA^{\ast}AA^{\ast}+A^{\ast}AA^{\ast}A.
\]
Now
\begin{eqnarray*}
AA^{\ast}A &=&
Q_{+}S^{\ast}Q_{-}\cdot Q_{-}SQ_{+}\cdot Q_{+}S^{\ast}Q_{-}
=Q_{+}S^{\ast}\cdot Q_{-}SQ_{+}S^{\ast}\cdot Q_{-} \\
&=& Q_{+}S^{\ast}\cdot SQ_{+}S^{\ast}\cdot Q_{-}=Q_{+}S^{\ast}Q_{-} \\
&=& A,
\end{eqnarray*}
hence
$A^{\ast}AA^{\ast}=A^{\ast}$
and
\[
(AA^{\ast}+A^{\ast}A)^{2}=AA^{\ast}+A^{\ast}A=:P,
\]
i.e.
$P$
is a selfadjoint projection and
$(A+A^{\ast})^{2}=P.$
Put
$A+A^{\ast}=:V.$
Then
$V=V^{\ast}$
and
$V^{2}=P$.
This implies
$V=E-F$
with selfadjoint projections
$E,\,F$,
where
$EF=0$
and
$E+F=PP.$

(ii) Conversely, assume
$J^{\ast}J=\EINS_{{\cal H}_{0}}+E-F.\,$
Then we have to prove
$D_{+}D_{-}=D_{-}D_{+}$,
or, equivalently,
$Q_{-}\cdot SQ_{+}S^{\ast}=SQ_{+}S^{\ast}\cdot Q_{-}.$
Put
$E+F=:P.$
We have
$A+A^{\ast}=E-F.$
Then
$(E-F)^{2}=E+F=P$,
i.e.
$(A+A^{\ast})^{2}=P$
or
$AA^{\ast}+A^{\ast}A=P.$
Put
$X:=AA^{\ast},\,Y:=A^{\ast}A$
Then
$X+Y=P$
and
$XY=0$.
This implies
$X^{2}=XP=PX$
and
$X^{2}(\EINS_{{\cal H}_{0}}-P)=(X(\EINS_{{\cal H}_{0}}-P))^{2}=0$,
hence
$X(\EINS_{{\cal H}_{0}}-P)=0$
or
$X=XP$
follows. Thus we get
\begin{equation}
X^{2}=X,
\end{equation}
i.e.
$X$
is a selfadjoint projection. Correspondingly, 
$Y$
is a selfadjoint projection, too. Recall that
\[
X=Q_{+}S^{\ast}Q_{-}\cdot Q_{-}SQ_{+}=Q_{+}S^{\ast}Q_{-}SQ_{+}.
\]
Then (18) yields
\[
Q_{+}S^{\ast}Q_{-}SQ_{+}S^{\ast}Q_{-}SQ_{+}=Q_{+}S^{\ast}Q_{-}SQ_{+},
\]
or, by multiplication with
$S^{\ast}Q_{-}S$
from the right,
\[
(Q_{+}\cdot S^{\ast}Q_{-}S)^{3}=(Q_{+}\cdot S^{\ast}Q_{-}S)^{2}.
\]
For brevity put
$Q_{+}S^{\ast}Q_{-}S=:B$.
Then
$(B^{2}-B)^{2}=0$
follows. This implies
$\vert B^{2}-B\vert=0$
and
$B^{2}=B$.
Therefore we obtain
\[
\mbox{s-lim}_{n\rightarrow\infty}(Q_{+}\cdot S^{\ast}Q_{-}S)^{n}=Q_{+}\cdot 
S^{\ast}Q_{-}S.
\]
Since the left hand side is a selfadjoint projection (onto the intersection subspace
$Q_{+}{\cal H}_{0}\cap S^{\ast}Q_{-}S{\cal H}_{0}$),
finally we get
$Q_{+}S^{\ast}Q_{-}S=S^{\ast}Q_{-}SQ_{+}$
or
\[
Q_{-}\cdot SQ_{+}S^{\ast}=SQ_{+}S^{\ast}\cdot Q_{-},
\]
and this is the assertion.

Now we prove the last statement. First we assume
$E=0$.
Then
$F=P$
and
\begin{equation}
J^{\ast}J=\EINS_{{\cal H}_{0}}-P.
\end{equation}
Then also
\[
JJ^{\ast}=D_{+}+D_{-}=W_{+}Q_{-}W_{+}^{\ast}+W_{-}Q_{+}W_{-}^{\ast}=
W_{+}(Q_{-}+SQ_{+}S^{\ast})W_{+}^{\ast}
\]
is a projection, i.e.
$Q_{-}+SQ_{+}S^{\ast}$
is a projection. This gives
$SQ_{+}S^{\ast}Q_{-}+Q_{-}SQ_{+}S^{\ast}=0$
But (19) implies
\[
Q_{+}S^{\ast}Q_{-}SQ_{+}+Q_{-}SQ_{+}S^{\ast}Q_{-}=-Q_{+}S^{\ast}Q_{-}-Q_{-}SQ_{+},
\]
hence
$Q_{-}SQ_{+}S^{\ast}Q_{-}=-Q_{-}SQ_{+}$
and
$Q_{-}SQ_{+}=0$
follows. Since
$P=-(Q_{+}S^{\ast}Q_{-}+Q_{-}SQ_{+}$,
we get
$P=F=0$.

On the other hand, if
$F=0$,
i.e.
$E=P$,
we have
$J^{\ast}J=\EINS_{{\cal H}_{0}}+P$
and
$P=Q_{+}S^{\ast}Q_{-}+Q_{-}SQ_{+}$.
Now, together with
$S$
also
$-S$
is an admissible scattering operator, assigned to a complete two-space scattering system
$\{\tilde{H},H_{0}\}$ (see [12, p. 238 ff.]). The corresponding identification
operator
$\tilde{J}$
satisfies
$\tilde{J}^{\ast}\tilde{J}=\EINS_{{\cal H}_{0}}-P$
and
$\tilde{J}\tilde{J}^{\ast}=\tilde{W}_{+}(Q_{-}+SQ_{+}S^{\ast})\tilde{W}^{\ast}_{+}$.
That is, also in this case
$Q_{-}+SQ_{+}S^{\ast}$
is a projection and we obtain, by similar arguments as before, that
$P=F=0.\,\Box$

\subsection{The Lax-Phillips semigroup }

As it is mentioned in Subsection 3.1 in the case
${\cal D}_{+}\bot{\cal D}_{-}$
Lax and Phillips introduced an important semigroup by a characteristic 
restriction of the LP-evolution.

In this Subsection we show that also in the case of commuting outgoing/incoming 
projections by an analogous restriction a semigroup can be introduced
which in the special case of mutually orthogonal outgoing and incoming
subspaces coincides with the LP-semigroup.

We start with the semigroup
\begin{equation}
D_{+}^{\bot}e^{-itH}D_{+}^{\bot}=D_{+}^{\bot}e^{-itH},\quad t\geq 0.
\end{equation}
Its transformation into the outgoing spectral representation yields the
characteristic
semigroup
$T_{+}(\cdot)$
(see Subsection 2.3). Now we define a second restriction of
(20)
by
\[
Z(t):=D_{+}^{\bot}e^{-itH}D_{-}^{\bot},\quad t\geq 0.
\]
A straightforward calculation gives
\[
Z(t)=W_{+}Q_{+}e^{-itH_{0}}SQ_{-}W_{-}^{\ast},
\]
i.e. the transformation into the outgoing spectral representation yields
\[
Z_{+}(t)=W_{+}^{\ast}Z(t)W_{+}=
Q_{+}e^{-itH_{0}}Q_{+}\cdot SQ_{-}S^{\ast}.
\]
Recall that the condition
$D_{+}D_{-}=D_{-}D_{+}$
is equivalent with (17).
Then we have

\vspace{3mm}

THEOREM 2. {\em If}
$D_{+}$
{\em and}
$D_{-}$
{\em commute then}
$Z_{+}(\cdot)$
{\em hence}
$Z(\cdot)$
{\em is a semigroup for}
$t\geq 0$.

\vspace{2mm}

Proof. We calculate
\begin{eqnarray*}
Z_{+}(t_{1})Z_{+}(t_{2}) &=&
Q_{+}e^{-it_{1}H_{0}}Q_{+}SQ_{-}S^{\ast}Q_{+}e^{-it_{2}H_{0}}Q_{+}SQ_{-}S^{\ast} \\
&=& Q_{+}e^{-it_{1}H_{0}}SQ_{-}S^{\ast}e^{-it_{2}H_{0}}SQ_{-}S^{\ast} \\
&=& Q_{+}Se^{-it_{1}H_{0}}Q_{-}e^{-it_{2}H_{0}}Q_{-}S^{\ast} \\
&=& Q_{+}Se^{-it_{1}H_{0}}e^{-it_{2}H_{0}}Q_{-}S^{\ast} \\
&=& Q_{+}e^{-i(t_{1}+t_{2})H_{0}}Q_{+}\cdot SQ_{-}S^{\ast} \\
&=& Z_{+}(t_{1}+t_{2}). \quad \Box
\end{eqnarray*} 

Note that 
$Q_{+}\cdot SQ_{-}S^{\ast}$
is the projection of the subspace
$Q_{+}{\cal H}_{0}\cap SQ_{-}{\cal H}_{0}$
hence we obtain
\[
Q_{+}SQ_{-}S^{\ast}{\cal H}_{0}=
Q_{+}{\cal H}_{0}\cap SQ_{-}{\cal H}_{0}=
{\cal H}^{2}_{+}\cap S{\cal H}^{2}_{-}=
{\cal H}^{2}_{+}\cap S({\cal H}^{2}_{+})^{\bot}=
{\cal H}^{2}_{+}\cap (S{\cal H}^{2}_{+})^{\bot}.
\]
This means: the elements of this subspace are exactly those vectors
$f\in{\cal H}^{2}_{+}$
which are orthogonal w.r.t.
$S{\cal H}^{2}_{+}$,
i.e.
$f\bot S{\cal H}^{2}_{+}$.

According to Theorem 2 this subspace is invariant w.r.t. the semigroup
$Z_{+}(\cdot)$. Moreover the semigroup vanishes on the orthogonal complement.
The restriction
\begin{equation}
Z_{+}(t)\restriction{\cal H}^{2}_{+}\cap(S{\cal H}^{2}_{+})^{\bot},\quad 
t\geq 0
\end{equation}
is a strongly continuous contractive semigroup which is a restriction of the 
characteristic semigroup
$T_{+}(\cdot)\restriction{\cal H}^{2}_{+}$
considered in Subsection 2.3.
This restriction we call the {\em generalized Lax-Phillips semigroup}.

\vspace{3mm}

REMARK 1. If even
$D_{+}D_{-}=0$,
i.e.
${\cal D}_{+}$
and
${\cal D}_{-}$
are orthogonal then Proposition 8 yields
$SQ_{+}=Q_{+}SQ_{+}.$
This means
$S{\cal H}^{2}_{+}\subseteq {\cal H}^{2}_{+}.$
In this case we obtain
\[
{\cal H}^{2}_{+}\cap(S{\cal H}^{2}_{+})^{\bot}={\cal H}^{2}_{+}\ominus
 S{\cal H}^{2}_{+},
\]
i.e. in this case
$Z_{+}(\cdot)$
acts on
${\cal H}^{2}_{+}\ominus S{\cal H}^{2}_{+}$
and it is nothing else than the original Lax-Phillips semigroup.Further it turns out that 
in this case
$S(\cdot)$
is holomorphic in
$\Bbb{C}_{+}$
with
$\sup_{z\in\Bbb{C}_{+}}\Vert S(z)\Vert\leq 1$
such that
$S(\lambda)=\mbox{s-lim}_{\epsilon\rightarrow+0}S(\lambda+i\epsilon)$.
That is, in this case the existence of the Lax-Phillips semigroup is simultaneously
coupled with strong implications on the analytic continuability of the scattering
matrix.

\vspace{3mm}

Next we study the spectral theory of (21). It
is a restriction of the characteristic semigroup 
$T_{+}(\cdot)\restriction{\cal H}^{2}_{+}$
whose spectral theory
is already known. Therefore, in view of the problem to characterize the eigenvalue 
spectrum of
(21)
the crucial question is: Which eigenvalues of the characteristic semigroup, i.e. of
$T_{+}(\cdot)$
on
${\cal H}^{2}_{+}$,
{\em survive} the restriction to the subspace
${\cal H}^{2}_{+}\cap(S{\cal H}^{2}_{+})^{\bot}$?
That is, for 
$f_{\zeta,k}\in{\cal N}_{\overline{\zeta}},\,\zeta\in\Bbb{C}_{-}$,
i.e.
\[
f_{\zeta,k}(\lambda):=\frac{k}{\lambda-\zeta},\quad 0\neq k\in{\cal K},
\]
one has to analyze the condition
$f_{\zeta,k}\bot S{\cal H}^{2}_{+}$
or, equivalently,
\begin{equation}
S^{\ast}f_{\zeta,k}\in{\cal H}^{2}_{-}.
\end{equation}
We have
\[
(S^{\ast}f_{\zeta,k})(\lambda)=S(\lambda)^{\ast}f_{\zeta,k}(\lambda)=
\frac{S(\lambda)^{\ast}k}{\lambda-\zeta}.
\]
Therefore (22) is equivalent to
\[
\int_{-\infty}^{\infty}\frac{S(\lambda)^{\ast}k}
{(\lambda-\zeta)(\lambda-z)}d\lambda=0,\quad z\in\Bbb{C}_{+},
\]
because of (3). In particular, (22) implies that
$(S^{\ast}f_{\zeta,k})(\cdot)$
has a holomorphic continuation into
$\Bbb{C}_{-}$.
Then
\begin{equation}
\Vert (S^{\ast}f_{\zeta,k})(z)\Vert_{\cal K}\leq\frac{\Vert k\Vert}{
\vert\mbox{Im}\,\zeta\vert},\quad z\in\Bbb{C}_{-},
\end{equation}
follows. On the other hand,
$\Bbb{C}_{-}\ni z\rightarrow (z-\zeta)(S^{\ast}f_{\zeta,k})(z)$
is the holomorphic continuation of
$\Bbb{R}\ni\lambda\rightarrow S(\lambda)^{\ast}k$
into
$\Bbb{C}_{-}$
and
$\zeta$
is a zero of this function. This implies
\[
\vert z-\zeta\vert\cdot\Vert(S^{\ast}f_{\zeta,k})(z)\Vert_{\cal K}\leq
\mbox{sup}_{\lambda\in\Bbb{R}}\Vert S(\lambda)^{\ast}k\Vert=\Vert k\Vert
\]
or
\begin{equation}
\Vert(S^{\ast}f_{\zeta,k})(z)\Vert_{\cal K}\leq
\frac{\Vert k\Vert}{\vert z-\zeta\vert},\quad \zeta\neq z\in\Bbb{C}_{-}.
\end{equation}
Therefore we obtain

\vspace{3mm}

PROPOSITION 9. {\em Let}
$(S^{\ast}f_{\zeta,k})(\cdot)$
{\em be holomorphic continuable into}
$\Bbb{C}_{-}$.
{\em Then}
$S^{\ast}f_{\zeta,k}\in{\cal H}^{2}_{-}$
{\em follows, i.e. the condition of holomorphic continuability of}
$S^{\ast}f_{\zeta,k}(\cdot)$
{\em into}
$\Bbb{C}_{-}$
{\em is sufficient for} (22).

\vspace{2mm}

Proof. Choose a square
$\Bbb{C}_{-}\supset G_{\epsilon}:=
\{z:\vert\mbox{Re}\,z-\mbox{Re}\,\zeta\vert\leq\epsilon,\,
\vert\mbox{Im}\,z-\mbox{Im}\,\zeta\vert\leq\epsilon\},\,\epsilon>0,$
and let
$y>0.$
If
$(\Bbb{R}-iy)\cap G_{\epsilon}=\emptyset$
then
\[
\int_{-\infty}^{\infty}\Vert(S^{\ast}f_{\zeta,k})(x-iy)\Vert^{2}_{\cal K}dx\leq
\Vert k\Vert^{2}\frac{\pi}{\epsilon},
\]
where we have used (24). If
$(\Bbb{R}-iy)\cap G_{\epsilon}\neq\emptyset$
then
\[
\int_{-\infty}^{\infty}=\int_{-\infty}^{\mbox{Re}\,\zeta-\epsilon}
+\int_{\mbox{Re}\,\zeta-\epsilon}^{\mbox{Re}\,\zeta+\epsilon}+
\int_{\mbox{Re}\,\zeta+\epsilon}^{\infty}.
\]
To estimate the first and the third term we use (24), for the second term we use (23). 
Thus in this case we obtain
\[
\int_{-\infty}^{\infty}\Vert(S^{\ast}f_{\zeta,k})(x-iy)\Vert^{2}_{\cal K}\leq
\Vert k\Vert^{2}\left(\frac{2}{\epsilon}+\frac{2\epsilon}{
\vert\mbox{Im}\,\zeta\vert^{2}}\right),
\]
i.e.
$\sup_{y>0}\int_{-\infty}^{\infty}\Vert(S^{\ast}f_{\zeta,k})
(x-iy)\Vert^{2}_{\cal K}dx<\infty.$
Therefore, according to the Paley-Wiener theorem, the assertion follows.
$\quad \Box$

\vspace{2mm}

REMARK 2. (i) Note that if
$(S^{\ast}k)(\cdot)$
is holomorphic continuable into
$\Bbb{C}_{-}$
and
$(S^{\ast}k)(\zeta)=0$
then
$(S^{\ast}f_{\zeta,k})(\cdot)$
is holomorphic continuable into
$\Bbb{C}_{-}$.

(ii) In the case
${\cal D}_{+}\bot{\cal D}_{-}$
the operator function
$S(\cdot)^{-1}$
is a priori holomorphic in
$\Bbb{C}_{-}.$
Then
$S^{\ast}f_{\zeta,k}(\cdot)$
is holomorphic in
$\Bbb{C}_{-}$
iff
$S(\zeta)^{-1}k=0.$
But this means that
$S(\cdot)$,
which is also analytically continuable int
$\Bbb{C}_{-}$,
has necessarily a pole at
$\zeta$
(see Lax and Phillips [1]).

\section{Acknowledgements}

It is a pleasure to thank Professors A. Bohm and M. Gadella for
discussions on resonances and their mathematical
description
at the conference on "Irreversible
Quantum Dynamics" in Trieste, 29th July - 2th August 2002,
Professor A. Bohm at the CFIF-Workshop on "Time
Asymmetric Quantum Theory: The Theory of Resonances", 23th -
26th July 2003, Lisbon and Professor Y. Strauss for discussions on
the subject at the 25th International Colloquium on Group Theoretical
Methods in Physics in Cocoyoc, Mexico, 2th-6th August 2004. 

\section{References}

\noindent [1] Lax, P., Phillips, R.: Scattering Theory,
Academic Press, New York London 1967

\vspace{3mm}

\noindent [2] Strauss, Y.: Sz.-Nagy-Foias Theory and Lax-Phillips Type
Semigroups in the Description of Quantum Mechanical Resonances,
mp-arc archive, no. 04-253

\vspace{3mm}

\noindent [3] Strauss, Y.: Resonances in the Rigged Hilbert
Space and L-P Scattering Theory, Internat. J. of Theor. Phys.
42, No. 10, 2285-2317 (2003) 

\vspace{0.3cm}

\noindent [4] Flesia, C. and Piron, C.: La the´orie de la diffusion de
Lax-Phillips dans le cas quantique, Helv. Phys. Acta 57, 697-703 (1984)

\vspace{3mm}

\noindent [5] Horwitz, L.P. and Piron, C.: The unstable system and irreversible
motion in quantum theory, Helvetica Physica Acta 66, 693-711 (1993)

\vspace{3mm}

\noindent [6] Eisenberg, E. and Horwitz, L.P.: Time irreversibility and
unstable systems in quantum physics, in
{\em Advances in Chemical Physics}, edited by I. Prigogine and S. Rice,
Vol. 99, Wiley, New York, pp. 245-297 (1997)

\vspace{3mm}

\noindent [7] Strauss, Y., Horwitz, L.P. and Eisenberg, E.: Representation
of quantum mechanical resonances in Lax-Phillips Hilbert space,
J. Math. Phys. 41, No. 12,8050-8071 (2000); DOI 10.1063/1.1310359

\vspace{3mm}

\noindent [8] Halmos, P.R.: Two subspaces, Trans. Amer. Math. Soc. 
144, 381-389 (1969)

\vspace{3mm}

\noindent [9] Kato, T.: Perturbation Theory for Linear Operators,
Springer Verlag Berlin 1976

\vspace{3mm}

\noindent [10] Baumg\"artel, H.: Gamov vectors for resonances, a Lax-Phillips
point of view, preprint, arXiv: math-ph/0407059

\vspace{3mm}

\noindent [11] Sz.-Nagy, B. and Foias, C.: {\em Harmonic Analysis of Operators
on Hilbert space}, North Holland Publishing Company, Amsterdam and London
(1970)

\vspace{0.3cm}

\noindent [12] Baumg\"artel, H. and Wollenberg, M.: Mathematical
Scattering Theory, Birkh\"auser, Basel Boston Stuttgart 1983

\vspace{0.3cm}

\noindent [13] Baumg\"artel, H.: Introduction to Hardy spaces.
Internat. J. of Theor. Phys., 42, No. 10, 2211-2221 (2003)

\vspace{0.3cm}

\noindent [14] Sinai, Ja. G.: Dynamical systems with multiple Lebesgue
spectrum, Izv. Akad. Nauk SSSR 25, 899-924 (1961), in Russian

\vspace{3mm}

\noindent [15] Achieser, N.I., Glasman, I. M.: Theorie der linearen Operatoren
im Hilbert-Raum, 8. erweiterte Auflage, Akademie-Verlag 1981

\end{document}